# Electron Pairing of Interfering Interface-Based Edge Modes

Sourav Biswas[*], Hemanta Kumar Kundu[*], Vladimir Umansky and Moty Heiblum

*Braun Center for Submicron Research, Department of Condensed Matter Physics,*

*Weizmann Institute of Science, Rehovot 7610001, Israel*

\* Equal contribution

Corresponding author: moty.heiblum@weizmann.ac.il

**The remarkable Cooper-like pairing phenomenon in the Aharonov-Bohm interference of a Fabry-Perot interferometer (FPI) – operating in the integer quantum Hall regime – remains baffling. Here, we report the interference of paired electrons employing 'interface edge modes'. These modes are born at the interface between the bulk of the FPI and an outer gated region tuned to a lower filling factor. Such configuration allows toggling the spin and the orbital of the Landau level (LL) of edge modes at the interface. We find that electron pairing occurs *only* when the two modes (the interfering outer and the first inner) belong to the *same* spinless LL.**

Despite extensive experimental and theoretical work since its discovery in the 80s', the quantum Hall effect (QHE) still provides a playground for extensive studies [1-5]. With an insulating bulk, transport studies are performed with the gapless edge modes. Due to 'bulk-edge' correspondence, the nature of the bulk's quantum state is revealed [6-8]. Among the ubiquitous studies, the important ones are: shot noise [9,10], interference and braiding [11-13], and thermal conductance measurements [14]. One of the intriguing and unexplained features is a pairing of electrons in bulk fillings $\nu_b > 2$. This phenomenon was observed in the interference of the outermost edge mode in a Fabry-Perot interferometer (FPI), with Aharonov-Bohm (AB) flux periodicity of $h/2e$, where $e$ is the electron charge, and $h$ is the Planck constant; suggesting interference of paired electrons ($e^* = 2e$) [15-17]. Efforts to understand this phenomenon failed thus far [18-20]. In order to understand the process leading to the pairing of electrons in a single edge mode, we utilize a new approach by interfering 'interface edge modes' [21,22].

Normal edge modes are confined to the interface between the plane of the two-dimensional electron gas (2DEG) and the 'vacuum.' The transverse Hall conductance $\sigma_{xy}$ is determined by the bulk filling $\nu_b$: $\sigma_{xy} = \frac{\nu_b e^2}{h}$. Interface modes, on the contrary, are confined to the interface between two bulk regions with fillings: $\nu_b$ and $\nu_g$ (the latter is a gated bulk). Consequently, the transverse interface mode conductance is, $\sigma_{xy} = \frac{\nu_{int} e^2}{h}$, with $\nu_{int} = (\nu_b - \nu_g)$, the interface filling. Recent charge and thermal transport measurements [22-25] reveal the potential of novel experiments employing 1D interface chiral modes. The schematic of the interface edge configuration, with $\nu_b = 2$ and $\nu_g = 1$, is shown in Fig. 1(a). At the interface, the counter-propagating modes 'gap' each other, leaving the inner mode of $\nu_b = 2$ at the interface, with $\nu_{int} = 1$. Two examples of gate-dependence of interface resistance with $\nu_b = 2$ & $4$ are shown in Fig. 1(b).

It is worth recalling the main observations in the past 'pairing experiments' [15,16] before we delve into new findings. Utilizing ubiquitous QHE integer edge modes, we stumbled on electron pairing while interfering the outermost edge mode belonging to the first Landau level (LL), LL1 ↓, in bulk fillings exceeding $\nu_b = 3$. The interfering mode was accompanied by the first inner mode LL1 ↑ and the second inner mode LL2 ↓, both unpartitioned and encircling inside the interferometer (the arrow denotes the spin). Surprisingly, the first inner edge mode, LL1 ↑, was found to control the coherence and determine the flux periodicity of the interfering outermost edge mode. Moreover, the second inner mode, LL2 ↓, had to be populated to observe pairing.

Here, we replace the ubiquitous edge modes with interface modes, thus allowing controlling the character of the two outermost modes involved in the interference process. We test an AB-FPI with bulk fillings, $2 \leq \nu_b \leq 6$, with different interface fillings determined by a neighboring gated bulk, thus adding crucial information important to the underlying pairing mechanism [Fig. 3 and Table].

The AB phase evolution is given by $\varphi_{AB} = \frac{2\pi BA}{\Phi_0}$, where $B$ is the applied magnetic field, $A$ is the area defined by the interfering edge mode, and $\Phi_0 = h/e$ the flux quantum [26]. With changing of the confined flux, the AB phase evolves as $\delta\varphi_{AB} = \frac{2\pi(B\delta A + A\delta B)}{\Phi_0}$, with the area changes by the modulation-gate voltage, $V_{MG}$, with $\delta A = \alpha \delta V_{MG}$, and $\alpha$ is proportional to the gate-2DEG capacitance. Assuming first-order interference, i.e., weak backscattering by the two quantum point contacts (QPCs), the interference is proportionate to $\cos \delta\varphi_{AB}$. Customarily, a 2D *pyjama* plot [27-29] in the $B - V_{MG}$ plane leads to periodicities, $\frac{1}{\Delta B} = \frac{A}{\Phi_0}$ and $\frac{1}{\Delta V_{MG}} = \frac{\alpha B}{\Phi_0}$.

Our interface edge-based FPI, with an internal lithographic area of $14.2\mu m^2$ [Fig. 2(a)], is fabricated in GaAs-AlGaAs heterostructure harboring high mobility 2DEG, with a 2D electron density of $1.7 \times 10^{11} cm^{-2}$, located 83nm below the surface. Hafnium oxide isolates different metallic contacts. An interior small grounded ohmic contact reduces the charging energy of the FPI and enables AB interference. AC voltage (1µV at 900kHz) is applied to the source, and the drain signal is amplified by a cold (1.5K) amplifier (with an LC circuit at its input), cascaded by a room-temperature amplifier. Measurements are performed at 10mK base temperature.

We first repeat the 'pairing experiments' with trivial edge modes. Starting at $\nu_{int} = \nu_b - \nu_g = 2 - 0 = 2$, the FPI's QPCs are tuned to partition the outermost edge mode and fully reflect the inner mode [Fig. 2(a)]. High visibility conductance oscillations with magnetic field and modulation gate voltage, characteristics of an FPI in a coherent AB regime, are observed [Fig. 2(b)]. The obtained periodicities in $B$ and $V_{MG}$ correspond to an area $\frac{\Phi_0}{\Delta B} = 11.3 \mu m^2$ and $\frac{1}{\Delta V_{MG}} = 109 V^{-1}$, respectively. These data ensure electron interference ($e^* = e$) with the flux quantum periodicity. The smaller AB area than the lithographic one indicates $\approx 200$nm depletion at the gate interface.

Interfering the outermost mode at the $\nu_{int} = 3 - 0 = 3$ configuration (see Fig. 3(a), and see Supplementary Fig. S3 for a detailed schematic) with QPCs' transmission $t \approx 0.88$ leads to frequencies $\frac{\Phi_0}{\Delta B} = 22.5 \mu m^2$ and $\frac{1}{\Delta V_{MG}} = 146.5 V^{-1}$ [Fig. 3(b)]; namely, $h/2e$ flux-periodicity and thus an apparent interfering charge $e^* = 2e$. As observed before [15], the filling of LL2 ↓ (second inner mode) is necessary to observe pairing in the interfering (outermost) mode in LL1 ↓.

To test the relation between the two outer modes, we toggled the interface mode's LL (spin and orbital) by tuning the $\nu_b$ and $\nu_g$; here, $\nu_{int} = 4 - 1 = 3$ [Fig. 3(a)] with LL1 ↓ gapped out. The interfering outermost mode belongs to the spin-split Landau level LL1 ↑ (with $t \approx 0.92$), and adjacent, the enclosed first inner, belongs to the orbital LL2 ↓. Note that a 'protective mode' - the second inner mode, belonging to the LL2 ↑ Landau level, circulates inside the FPI. The observed periodicities, $\frac{\Phi_0}{\Delta B} = 11.4 \mu m^2$ and $\frac{1}{\Delta V_{MG}} = 70.8 V^{-1}$ [Fig. 3(c)], are similar to those of $2 - 0$ configuration, namely, no pairing. See Supplementary Fig. S4 for data with a strongly pinched QPC.

What is the main difference between the $3 - 0$ and $4 - 1$ configurations? In the $3 - 0$ case, the outermost interfering mode and the first-inner mode belong to the same spinless Landau level, LL1, i.e., they share the same orbital (but carry opposite spins). In the $4 - 1$ case, though the pairs are spinless, the interfering mode belongs to LL1, while the first-inner mode belongs to LL2. As seen above, pairing occurs when the two outer modes belong to the same spinless Landau level. Indeed, pairing is also observed in the $5 - 2$ configuration; i.e., the outermost mode belongs to LL2 $\downarrow$, the first-inner mode belongs to LL2 $\uparrow$, and the protective mode is LL3 $\downarrow$ (see Supplementary Fig. S5).

To further confirm the robustness and universality of the above pairing, we interfere various modes (inner and outer) at the bulk fillings $4 \leq \nu_b \leq 6$ (Supplementary Figs. S6 & S7). The obtained results are tabulated in the Table. The different realizations allow a clear view of the needed conditions to observe pairing.

Aharonov-Bohm interference of paired electrons in integer quantum Hall remains a puzzling observation without an explanation. Together with past data, our new results, based on interfering interface edge mode, show: (a) Pairing occurs between modes belonging to the same spinful Landau level, hence, the pairs are spinless. (b) The paired modes must be accompanied by an inner mode (belonging to a higher Landau level). This mode does not affect the pairing but seems to protect (screen) the paired modes from the bulk. We did not observe yet such an effect in the fractional regime.

We thank D. E. Feldman for the useful discussion. We acknowledge the continuous support of the Sub-Micron Center staff. M.H. acknowledges the support of the European Research Council under the European Community's Seventh Framework Program (FP7/2007-2013)/ERC under grant agreement number 713351, the partial support of the Minerva Foundation with funding from the Federal German Ministry for Education and Research, under grant number 713534.

**References:**

[1]     K. von Klitzing, T. Chakraborty, P. Kim, V. Madhavan, X. Dai, and J. McIver, 40 years of the quantum Hall effect, Nature Review Physics **2**, 397 (2020).

[2]     B. I. Halperin and J. K. Jain, *Fractional Quantum Hall Effects: New Developments* (WORLD SCIENTIFIC, 2020), Fractional Quantum Hall Effects.

[3]     M. Z. Hasan and C. L. Kane, Colloquium: Topological insulators, Reviews of Modern Physics **82**, 3045 (2010).

[4]     S. D. Sarma and A. Pinczuk, *Perspective in Quantum Hall Effects* (Wiley, 1996), Perspective in Quantum Hall Effects.

[5]     R. Prange and S. M. Girvin, *The Quantum Hall Effect* (Springer Verlag, 1990).

[6]     B. I. Halperin, Quantized Hall conductance, current-carrying edge states, and the existence of extended states in a two-dimensional disordered potential, Physical Review B **25**, 2185 (1982).

[7]     X. G. Wen, Gapless boundary excitations in the quantum Hall states and in the chiral spin states, Physical Review B **43**, 11025 (1991).

[8]     M. Heiblum and D. E. Feldman, Edge probes of topological order, International Journal of Modern Physics A **35** 2030009 (2020).

[9]     R. dePicciotto, M. Reznikov, M. Heiblum, V. Umansky, G. Bunin, and D. Mahalu, Direct observation of a fractional charge, Nature **389**, 162 (1997).

[10]    M. Heiblum, Quantum shot noise in edge channels, Physica Status Solidi B-Basic Solid State Physics **243**, 3604 (2006).

[11]    Y. Ji, Y. Chung, D. Sprinzak, M. Heiblum, D. Mahalu, and H. Shtrikman, An electronic Mach-Zehnder interferometer, Nature **422**, 415 (2003).

[12]    J. Nakamura, S. Liang, G. C. Gardner, and M. J. Manfra, Direct observation of anyonic braiding statistics, Nature Physics **16**, 931 (2020).

[13]    H. K. Kundu, S. Biswas, N. Ofek, V. Y. Umansky, and M. Heiblum, Anyonic interference and braiding phase in a Mach-Zehnder Interferometer, Nature Physics (2023).

[14]    M. Banerjee, M. Heiblum, V. Umansky, D. E. Feldman, Y. Oreg, and A. Stern, Observation of half-integer thermal Hall conductance, Nature **559**, 205 (2018).

[15]    H. K. Choi, I. Sivan, A. Rosenblatt, M. Heiblum, V. Umansky, and D. Mahalu, Robust electron pairing in the integer quantum hall effect regime, Nature Communications **6**, 7435 (2015).

[16]    I. Sivan, R. Bhattacharyya, H. K. Choi, M. Heiblum, D. E. Feldman, D. Mahalu, and V. Umansky, Interaction-induced interference in the integer quantum Hall effect, Physical Review B **97**, 125405 (2018).

[17]  A. Demir, N. Staley, S. Aronson, S. Tomarken, K. West, K. Baldwin, L. Pfeiffer, and R. Ashoori, Correlated Double-Electron Additions at the Edge of a Two-Dimensional Electronic System, Physical Review Letters **126**, 256802 (2021).

[18]  G. A. Frigeri and B. Rosenow, Electron pairing in the quantum Hall regime due to neutralon exchange, Physical Review Research **2**, 043396 (2020).

[19]  G. A. Frigeri, D. D. Scherer, and B. Rosenow, Sub-periods and apparent pairing in integer quantum Hall interferometers, Europhysics Letters **126**, 67007 (2019).

[20]  D. Ferraro and E. Sukhorukov, Interaction effects in a multi-channel Fabry-Perot interferometer in the Aharonov-Bohm regime, SciPost Physics **3**, 014 (2017).

[21]  B. Dutta, W. Yang, R. Melcer, H. K. Kundu, M. Heiblum, V. Umansky, Y. Oreg, A. Stern, and D. Mross, Distinguishing between non-abelian topological orders in a quantum Hall system, Science **375**, 193 (2022).

[22]  S. Biswas, R. Bhattacharyya, H. K. Kundu, A. Das, M. Heiblum, V. Y. Umansky, M. Goldstein, and Y. Gefen, Shot noise does not always provide the quasiparticle charge, Nature Physics **18**, 1476 (2022).

[23]  B. Dutta, V. Umansky, M. Banerjee, and M. Heiblum, Isolated ballistic non-abelian interface channel, Science **377**, 1198 (2022).

[24]  M. Yutushui, A. Stern, and D. F. Mross, Identifying the $\nu=5/2$ Topological Order through Charge Transport Measurements, Physical Review Letters **128**, 016401 (2022).

[25]  C. Lin, M. Hashisaka, T. Akiho, K. Muraki, and T. Fujisawa, Time-resolved investigation of plasmon mode along interface channels in integer and fractional quantum Hall regimes, Physical Review B **104**, 125304 (2021).

[26]  B. I. Halperin, A. Stern, I. Neder, and B. Rosenow, Theory of the Fabry-Perot quantum Hall interferometer, Physical Review B **83**, 155440 (2011).

[27]  Y. M. Zhang, D. T. McClure, E. M. Levenson-Falk, C. M. Marcus, L. N. Pfeiffer, and K. W. West, Distinct signatures for Coulomb blockade and Aharonov-Bohm interference in electronic Fabry-Perot interferometers, Physical Review B **79**, 241304(R) (2009).

[28]  N. Ofek, A. Bid, M. Heiblum, A. Stern, V. Umansky, and D. Mahalu, Role of interactions in an electronic Fabry–Perot interferometer operating in the quantum Hall effect regime, Proceedings of the National Academy of Sciences **107**, 5276 (2010).

[29]   I. Sivan, H. K. Choi, J. Park, A. Rosenblatt, Y. Gefen, D. Mahalu, and V. Umansky, Observation of interaction-induced modulations of a quantum Hall liquid's area, Nature Communications **7**, 12184, 12184 (2016).

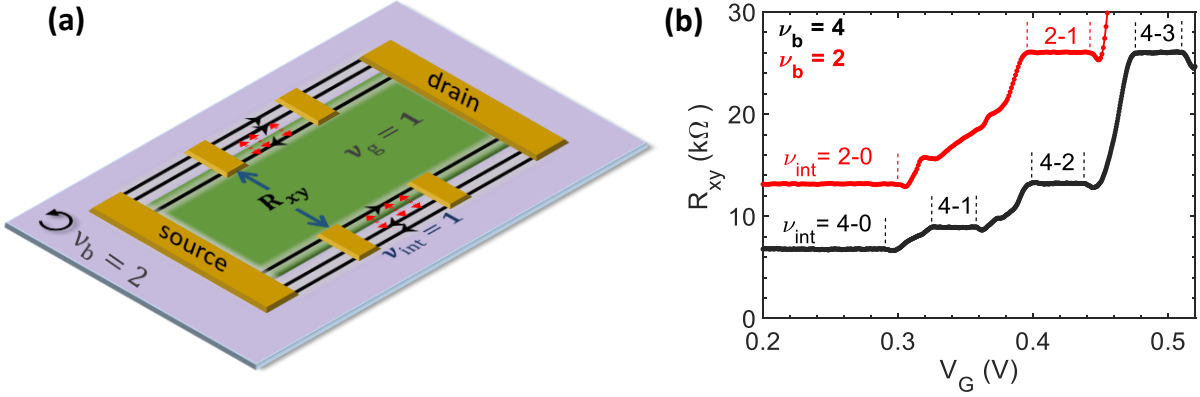

**Figure 1 | Interface edge modes. (a)** Schematic of interface edge modes. The bulk and gated region are shown in purple and green, respectively. The ohmic contacts (shown in yellow), made with Ni/Au/Ge evaporation followed by rapid thermal annealing, are placed at the interface. The gates are patterened by Pd/Au in cold evaporation. The bulk (gate) filling $\nu_b$ ($\nu_g$) is tuned by the magnetic field (gate voltage $V_G$). When $\nu_b = 2$ and $\nu_g = 1$, full equilibration (shown by red arrows) between counter-propagating modes leads to one mode ($\nu_{int} = 1$) left at the interface. **(b)** Interface Hall resistance showing various integer plateaus with gate voltage $V_G$ at the bulk filling $\nu_b = 2$ and 4. The bias cooling voltage for the gates is 0.55V. The full depletion of the gate with filling underneath $\nu_g = 0$ starts at $V_G \approx 0.3$V.

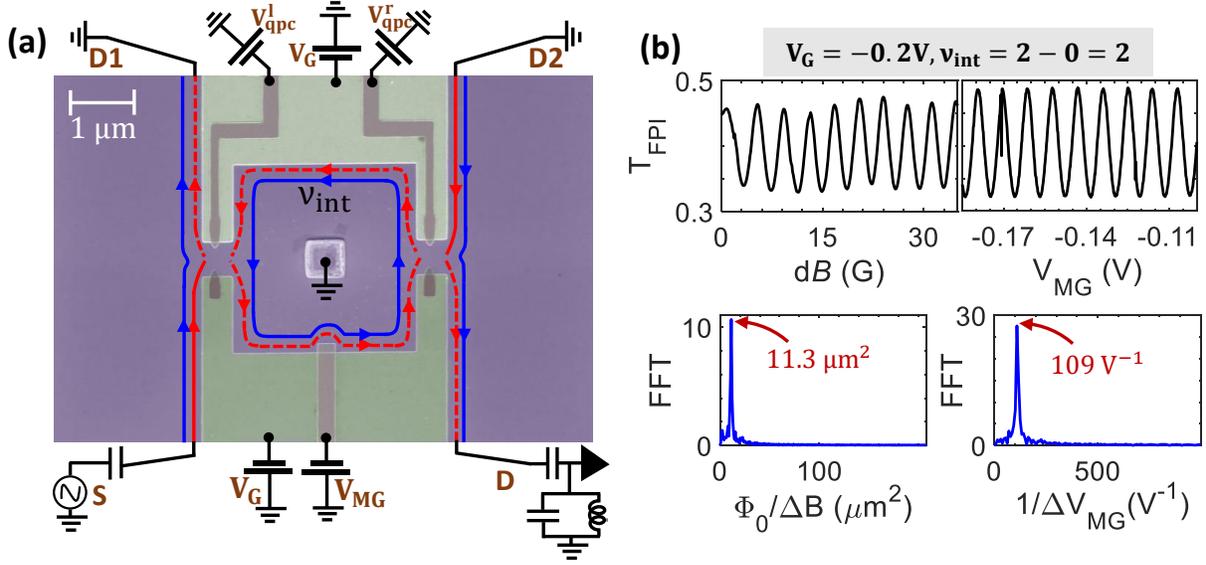

**Figure 2 | Interface mode-based Fabry-Perot interferometer and Aharonov-Bohm oscillations.** (a) False color scanning electron micrograph (SEM) image of the Fabry-Perot interferometer (FPI) with a grounded ohmic in the interior bulk. The lithographic internal area is 14.2μm². The bulk region is shown in purple. The top and bottom gates' (shown in green) depletion characteristics are identical, and the voltage $V_G$ tunes the filling underneath, hence the interface edge filling. The split gates as quantum point contacts, QPCs (brown) are separated from the lower gate (green) by 5nm Hafnium oxide. The incoming edge modes from the source (S) contact is biased with an AC voltage. The drain (D) contact measures the conductance as the transmission $T_{FPI}$ through the FPI. The modulation gate (300nm wide) sitting at the periphery of the interferometer tunes the area by $V_{MG}$. See Supplementary Fig. S1 for more details of the device. (b) Traces of $T_{FPI}$ with magnetic field and modulation gate voltage showing AB oscillations when the outer edge at $2 - 0 = 2$ is weakly partitioned. The average transmission is $\overline{T_{FPI}} \approx 0.4$, and the individual QPC transmission (for the interfering outer edge) is $t \approx 0.89$. We assume the left and right QPC are identical with the transmission probability $t_{qpc}^l = t_{qpc}^r = t$, and thus the $\overline{T_{FPI}} = |t|^2$. The first Fourier transformations (FFTs) with a single peak frequency are shown.

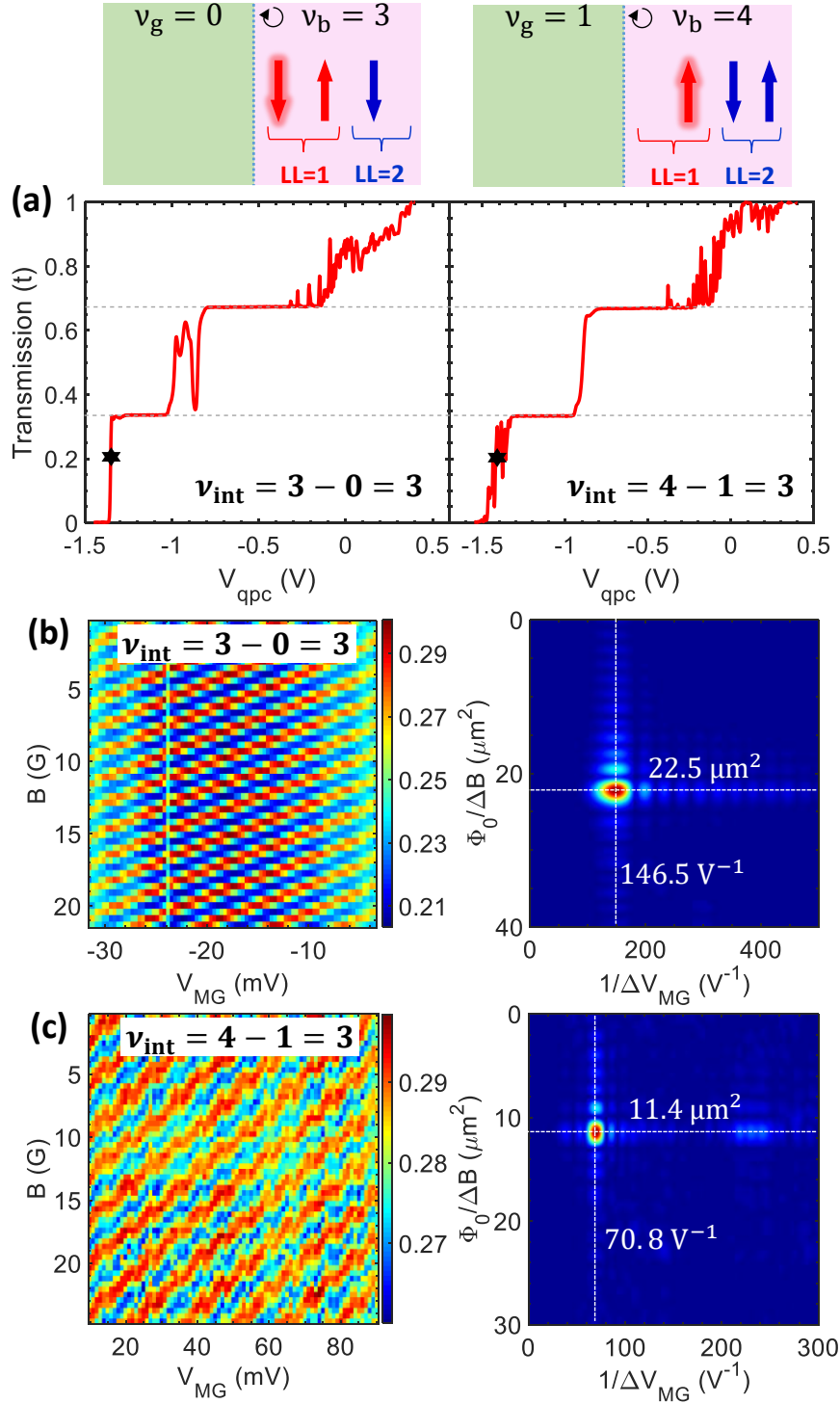

**Figure 3 | Outer edge interference for two types of engineered $\nu_{int} = 3$, namely, $3 - 0 = 3$ and $4 - 1 = 3$.** (a) The transmission ($t$) in a quantum point contact (QPC) as a function of the QPC-gate-voltage $V_{qpc}$ showing three plateaus for both conditions for $\nu_{int} = 3$. The star symbol represents a typical partitioning of the outer mode for interference. The schematics on top represent

the spin states (not to be confused with the chirality) of the edge modes (after full equilibration) that propagate for $3-0$ and $4-1$ edge configurations. The glowing one corresponds to the interfering (outer) edge. **(b)** The characteristic Aharonov-Bohm (AB) pyjama for $3-0$ (left) when the QPCs are set at $t \approx 88\%$ and the average transmission is $\overline{T_{\text{FPI}}} \approx 0.254$. The corresponding first Fourier transformation (FFT) with frequency values is shown on the right. **(c)** The AB pyjama and the FFT for $4-1$, when the QPC transmission is $t \approx 92\%$, and $\overline{T_{\text{FPI}}} \approx 0.28$. The values of $\frac{\Phi_0}{\Delta B}$ and $\frac{1}{\Delta V_{\text{MG}}}$ clearly show the interference of paired electrons ($2e$) at $3-0=3$, while the pairing is *not* observed at $4-1=3$. The comparisons are made with $2-0=2$ outer edge, see also the Table.

| B (Tesla) | Interference configuration | $\frac{\Phi_0}{\Delta B}$ ($\mu m^2$) | $\frac{1}{\Delta V_{MG}}$ ($V^{-1}$) | Flux-periodicity | Interfering charge ($e^*$) | Paring |
|---|---|---|---|---|---|---|
| 3.7 | $2 - 0 = 2$ <br> ↓↑ (outer) | 11.3 | 109 | $\Phi_0$ | $e$ | No |
| 2.45 | $3 - 0 = 3$ <br> ↓↑↓ (outer) | 22.5 | 146.5 | $\cong \frac{\Phi_0}{2}$ | $\cong 2.03e$ | Yes |
| 1.85 | $4 - 1 = 3$ <br> ↑↓↑ (outer) | 11.4 | 70.8 | $\cong \Phi_0$ | $\cong 1.3e$ | No |
| 1.45 | $5 - 2 = 3$ <br> ↓↑↓ (outer) | 19 | 104 | $\cong \frac{\Phi_0}{2}$ | $\cong 2.4e$ | Yes |
| 1.85 | $4 - 0 = 4$ <br> ↓↑↓↑ (2nd inner) | 7.5 | 64.77 | $\cong \Phi_0$ | $\cong 1.18e$ | No |
| 1.45 | $5 - 0 = 5$ <br> ↓↑↓↑↓ (2nd inner) | 12.4 | 118 | $\cong \frac{\Phi_0}{2}$ | $\cong 2.7e$ | Yes |
| 1.22 | $6 - 0 = 6$ <br> ↓↑↓↑↓↑ (2nd inner) | 19 | 87 | $\cong \frac{\Phi_0}{2}$ | $\cong 2.4e$ | Yes |
| 1.22 | $6 - 2 = 4$ <br> ↓↑↓↑ (outer) | 21 | 86 | $\cong \frac{\Phi_0}{2}$ | $\cong 2.4e$ | Yes |

**Table | Summary of the results for different interference configurations.** Obtained frequency values (in $B$ and $V_{MG}$ dependent oscillation traces, pyjamas), the estimated interfering charge ($e^*$), and the flux-periodicity ($\Delta\Phi$) for the interfering edge mode at interface filling $v_{int} = v_b - v_g$. The modes (spin states) are shown by colored arrows with the first one starting from the left being outer (nearest to the bulk-gate interface) and the next ones being 1st inner, 2nd inner, and so on. The interfering one is shown by the dashed arrow. The AB flux periodicity is obtained by comparing the field periodicities $\Delta B$ between filling factors and $v_{int} = 2 - 0$ outer edge. The less area ($\Phi_0/\Delta B$) for an inner edge is attributed to the presence of two more edge channels from the boundary. The interfering charge $e^*$ at $B = B_2$ is estimated using gate voltage periodicities $\Delta V_{MG}$ and the relation $\frac{e^*}{e} = \frac{B_1 \Delta V_{MG1}}{B_2 \Delta V_{MG2}}$ with $e^* = e$ (e.g., at $2 - 0 = 2$ outer edge) at $B = B_1$ [12,13]. Note

that the equation assumes the mutual capacitance α between the interfering edge channel and modulation gate to be constant. However, α varies with the number of fully transmitted edge channels [28]. Therefore, the comparison of an inner mode with the outer one using this relation is not ideal and provides quite a large error in the value of $e^*$.

# Supplementary Data

# "Electron Pairing of Interfering Interface-Based Edge Modes"

Sourav Biswas[*], Hemanta Kumar Kundu[*], Vladimir Umansky and Moty Heiblum

*Braun Center for Submicron Research, Department of Condensed Matter Physics,*

*Weizmann Institute of Science, Rehovot 7610001, Israel*

\* Equal contribution

Corresponding author: moty.heiblum@weizmann.ac.il

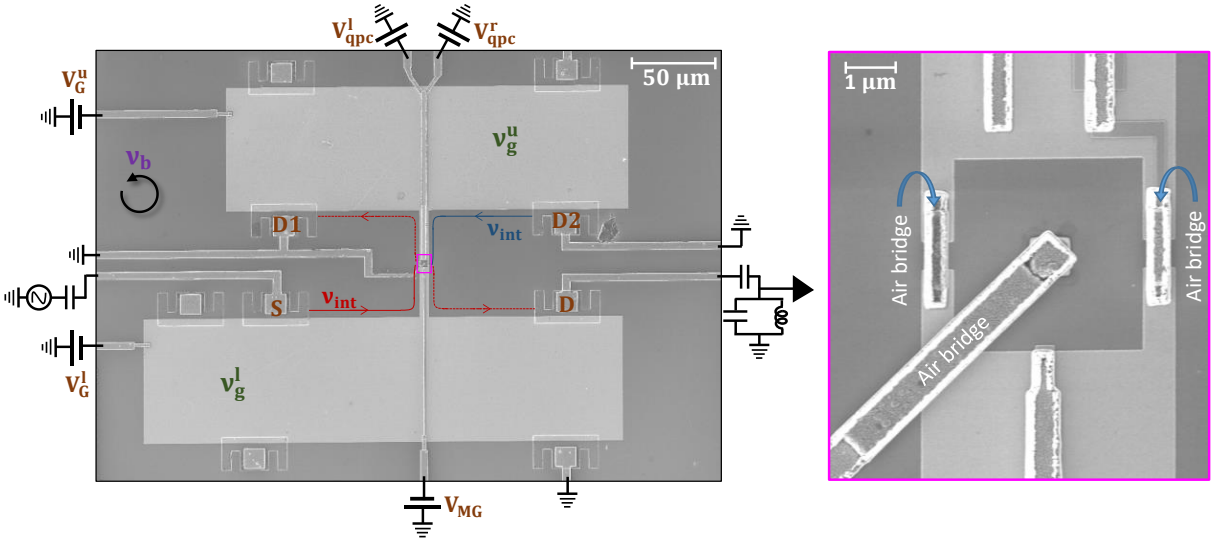

**Figure S1 | Large-scale SEM image of the device.** Left: SEM image of the whole device structure with various ohmic contacts, gates, and leads. Right: A zoomed image of the FPI showing the air bridge connections on the QPCs and the center ohmic contact. A partial broken lead from the left QPC is to be ignored.

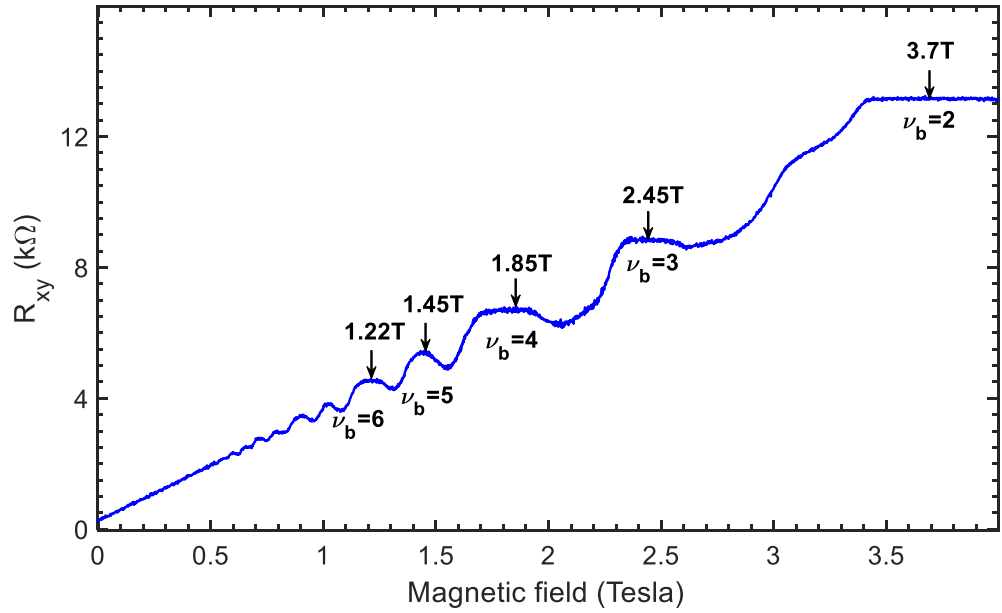

**Figure S2 | Hall measurement of the bulk.** Two terminal Hall resistance ($R_{xy}$) of the bulk as a function of the magnetic field ($B$) showing the QH plateaus. The bulk filling factors from $\nu_b = 2$ to 6 are marked with the respective $B$ value at which the measurements were performed.

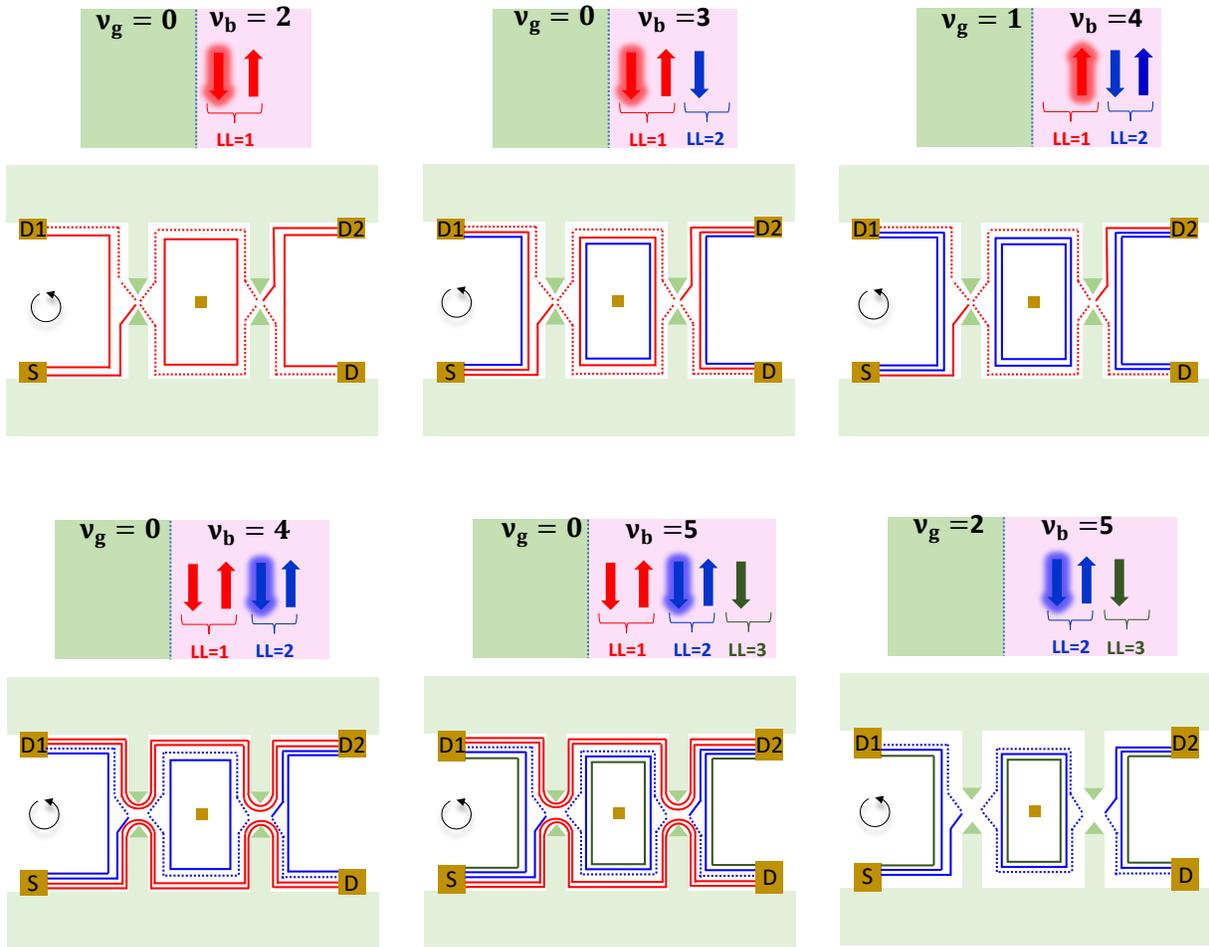

**Figure S3 | Edge-mode schematics for various interference configurations:** Schematics of the edge modes propagation around the gates, and inside the interferometer for different interference conditions. The respective LL (spin) modes with the interfering one glow are shown on top.

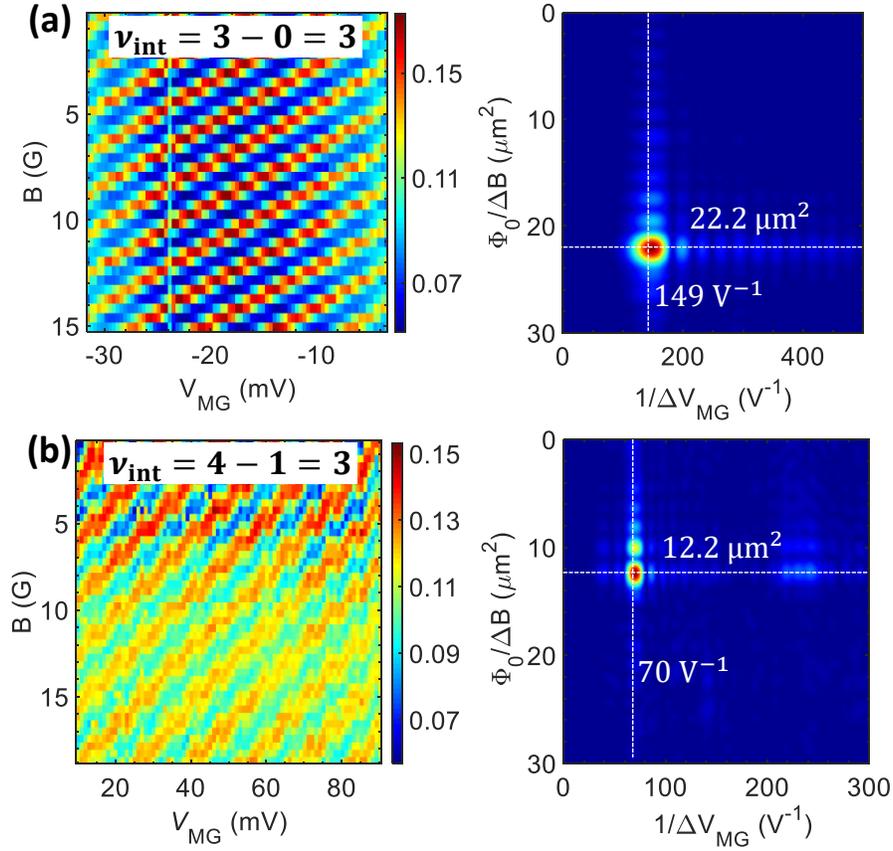

**Figure S4 | Outer edge interference for $\nu_{int} = 3$, when QPCs are relatively pinched.** The characteristic AB pyjamas and their FFT for **(a)** $3 - 0 = 3$ with pairing, when $t \approx 0.57$, and **(b)** $4 - 1 = 3$ without pairing, when $t \approx 0.57$.

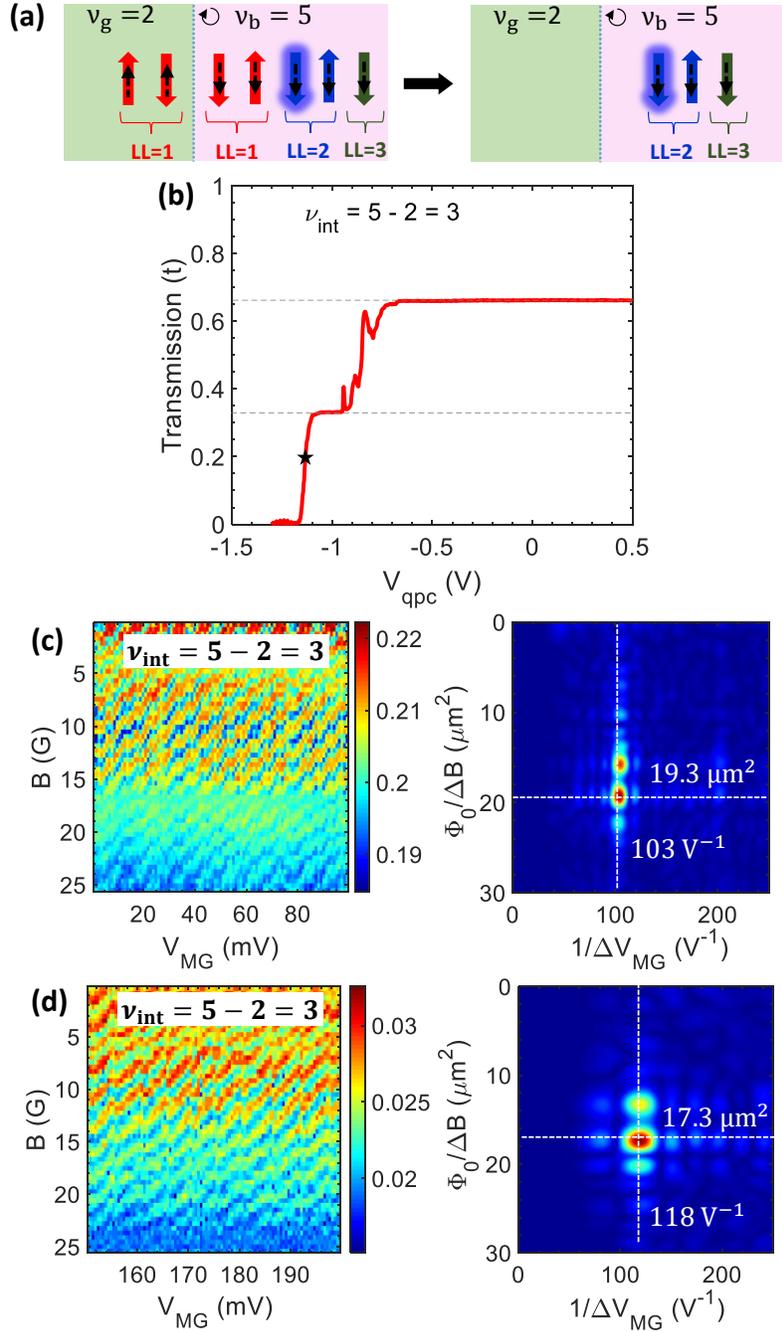

**Figure S5 | Results for $\nu_{\text{int}} = 5 - 2 = 3$**: **(a)** Schematics of the spin states (bold arrow) of the edge modes for $\nu_b = 5$ and $\nu_g = 2$, before and after the edge equilibration. The black dashed arrows represent the chirality of the edges. **(b)** The transmission of the QPC. The innermost LL3 ↓ edge is fully reflected when the $V_G$ is set for $\nu_g = 2$, possibly due to arrangements of modes after equilibration. **(c)** The AB pyjama when $t \approx 0.74$ and $\overline{T_{\text{FPI}}} \approx 0.183$, and the FFT. **(d)** The same with pinched QPC of $t \approx 0.27$ and $\overline{T_{\text{FPI}}} \approx 0.025$, and the FFT.

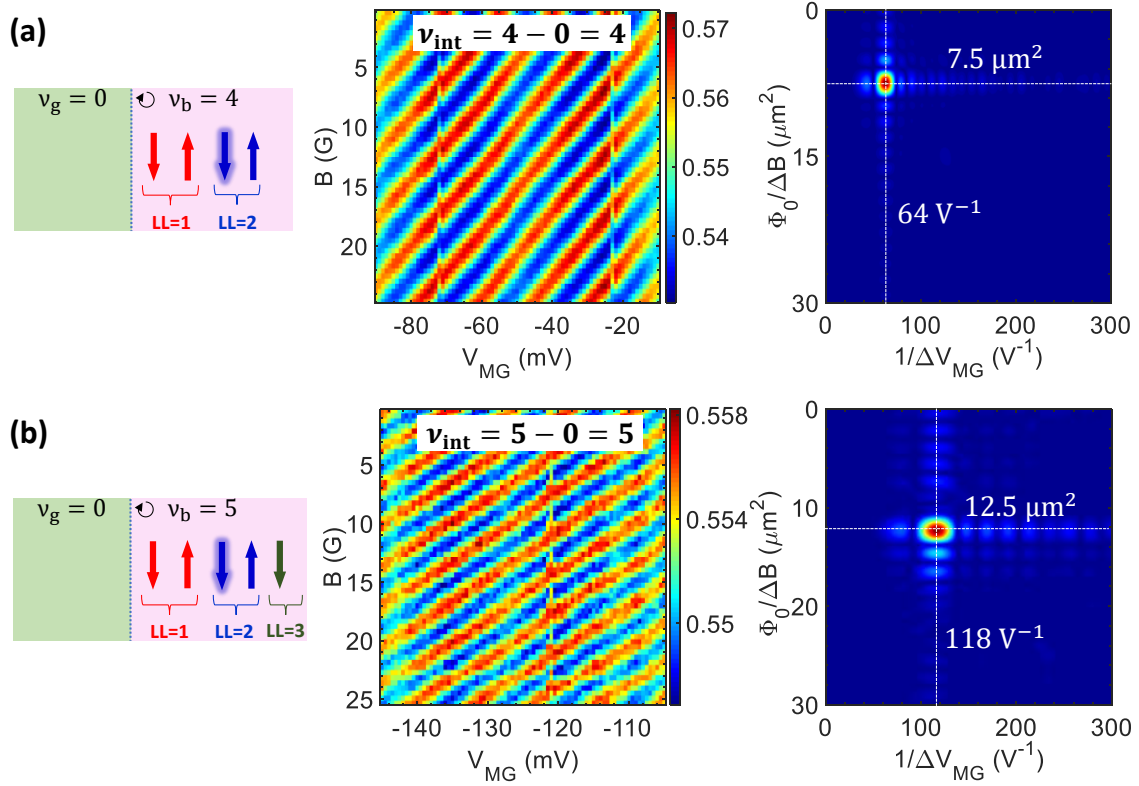

**Figure S6 | Interference of the inner edge of LL2 and the effect of edge screening to the bulk.** The schematics show the spin states of edge modes for the interface configurations of **(a)** $4-0$, and **(b)** $5-0$. The interfering inner edge mode (shown in the glown) corresponds to LL2. Respective Aharonov-Bohm pyjamas and their FFTs are shown on the right. The QPC transmissions are $t \approx 0.47$ and $t \approx 0.87$ for $4-0$ and $5-0$, respectively. The same LL2 edge with the same spin orientations shows the almost double frequency at $5-0$, compared to that at $4-0$ case. These results further strengthen the pairing conditions.

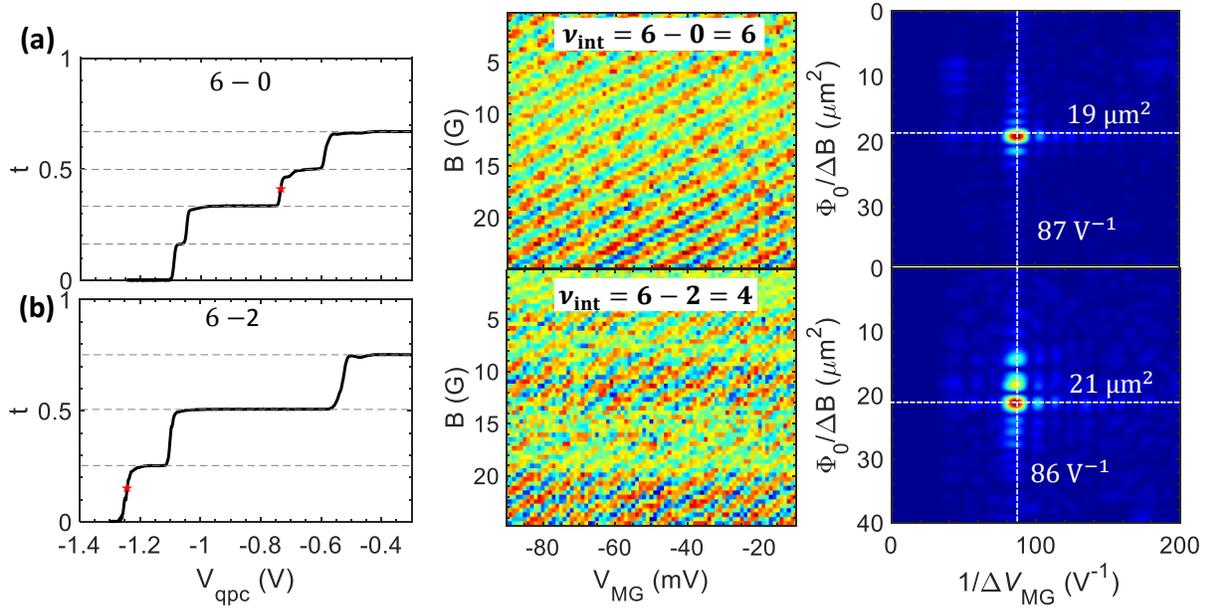

**Figure S7 | Results at bulk filling of $\nu_b = 6$**. The transmission in a QPC showing equispaced integer plateaus, and the interfering LL2 mode marked by the star symbol for **(a)** $6 - 0 = 6$, and **(b)** $6 - 2 = 4$ interface configurations. The AB interference patterns and the FFTs are shown on the right. For both cases, the frequencies show pairing, irrespective of whether two LL1 modes are fully gapped out or fully transmitted through the FPI.